\documentclass[aps,pra,twocolumn,superscriptaddress]{revtex4-2}

\usepackage{amsmath,amssymb,graphicx}
\usepackage{color,times}
\usepackage{siunitx}
\usepackage{xcolor}
\definecolor{darkblue}{rgb}{0, 0, 0.8}

\usepackage[colorlinks=true, breaklinks=true, linkcolor=darkblue, citecolor=darkblue, urlcolor=darkblue]{hyperref}

\newcommand{\doilink}[2]{\href{http://dx.doi.org/#1}{#2}}

\begin{document}

\title{Single Atoms with 6000-Second Trapping Lifetimes in Optical-Tweezer Arrays\\
at Cryogenic Temperatures}

\author{Kai-Niklas~Schymik} 
\affiliation{Universit\'e Paris-Saclay, Institut d'Optique Graduate School,\\
CNRS, Laboratoire Charles Fabry, 91127 Palaiseau Cedex, France}
\author{Sara~Pancaldi}  
\affiliation{Universit\'e Paris-Saclay, Institut d'Optique Graduate School,\\
CNRS, Laboratoire Charles Fabry, 91127 Palaiseau Cedex, France}
\author{Florence~Nogrette}
\affiliation{Universit\'e Paris-Saclay, Institut d'Optique Graduate School,\\
CNRS, Laboratoire Charles Fabry, 91127 Palaiseau Cedex, France}
\author{Daniel~Barredo}
\affiliation{Universit\'e Paris-Saclay, Institut d'Optique Graduate School,\\
CNRS, Laboratoire Charles Fabry, 91127 Palaiseau Cedex, France}
\affiliation{Nanomaterials and Nanotechnology Research Center (CINN-CSIC), Universidad de Oviedo (UO), Principado de Asturias, 33940 El Entrego, Spain}

\author{Julien~Paris}
\affiliation{My Cryo Firm, 20 villa des carri\`eres, 94120 Fontenay-sous-Bois, France}
\author{Antoine~Browaeys}
\affiliation{Universit\'e Paris-Saclay, Institut d'Optique Graduate School,\\
CNRS, Laboratoire Charles Fabry, 91127 Palaiseau Cedex, France}
\author{Thierry~Lahaye}
\affiliation{Universit\'e Paris-Saclay, Institut d'Optique Graduate School,\\
CNRS, Laboratoire Charles Fabry, 91127 Palaiseau Cedex, France}

\date{\today}

\begin{abstract}
We report on the trapping of single Rb atoms in tunable arrays of optical tweezers in a cryogenic environment at $\sim 4$~K. We describe the design and construction of the experimental apparatus, based on a custom-made, UHV compatible, closed-cycle cryostat with optical access. We demonstrate the trapping of single atoms in cryogenic arrays of optical tweezers, with lifetimes in excess of $\sim6000$~s, despite the fact that the vacuum system has not been baked out. These results open the way to large arrays of single atoms with extended coherence, for applications in large-scale quantum simulation of many-body systems, and more generally in quantum science and technology. 
\end{abstract}

\maketitle

For most applications of quantum science and technology, whatever the experimental platform, scaling up the number of individually-controlled quantum objects is a major subject of research, as this is a necessary condition for practical use~\cite{Alexeev2021}. Over the last few years, tweezers atom arrays have emerged  as a very versatile platform for quantum science, with applications ranging from quantum simulation of many-body systems~\cite{Browaeys2020} to quantum metrology~\cite{Madjarov2019,Norcia2019} and quantum computing~\cite{Saffman2016,Henriet2020}. Large arrays with up to $\sim200$ atoms are now used for quantum simulation of spin systems~\cite{Scholl2021,Ebadi2021}. They are assembled atom by atom, using a moving optical tweezers, from an initially disordered configuration. One of the current challenges in the field is to scale up the atom number, while preserving, or even increasing, the coherence of the system. 

A natural way to achieve this goal is to operate the tweezers arrays in a cryogenic environment at a temperature of a few Kelvin. A first beneficial effect is that the residual pressure is considerably smaller than at room temperature, which reduces collisions of the trapped atoms with the residual gas. This allows to increase the trapping lifetime of atoms in tweezers, which is one of the limiting factors in the assembly of large arrays, as the assembly time increases with system size. For a sequential assembly scheme, as used e.g. in \cite{Schymik2020}, increasing the trapping lifetime by a factor $\alpha$ allows for an increase in atom number by roughly $\sqrt{\alpha}$~\cite{NoteSQRT}. A second benefit is that black-body radiation (BBR), which scales as $T^4$, is considerably reduced in such an environment, making BBR-induced transitions between Rydberg levels almost negligible. For low angular momentum Rydberg states, this results in a typical increase of the Rydberg lifetime by a factor 2 to 3~\cite{Beterov2009}, with a direct impact on coherence and gate fidelities~\cite{Cong2021}. The inhibition of BBR-induced transitions would also be beneficial for Rydberg dressing experiments where they are a serious limitation \cite{Goldschmidt2016,Zeiher2016}. For circular states, the lifetime increases by several orders of magnitude in a cryogenic environment~\cite{Haroche2013}, motivating their use for quantum computing and simulation \cite{Xia2013,Nguyen2018,Cohen2021}. Finally, cryogenic single-atom trapping is also required, albeit at much lower temperatures, for coupling single atoms to microwave resonators in order to build hybrid systems~\cite{Pritchard2014}.

Here we demonstrate the trapping of single atoms in arrays of optical tweezers in a cryogenic environment at 4~K. We first describe the design, construction, and characterization of the setup, based on a closed-cycle cryostat where we use only UHV-compatible components. We then show how laser cooling and trapping of Rb atoms in the setup is obtained without any strong change as compared to a room-temperature setup.  We finally show that we can trap single atoms in arrays of tweezers, with measured lifetimes in the tweezers of over $\sim6000$~s, a 300-fold improvement compared to our current room-temperature setup. 

\section{Experimental apparatus}
\subsection{Cryostat design}

Adapting an atom tweezers setup for operating it at cryogenic temperatures comes with many specific technical constraints. This means that a straightforward use of the cryogenic solutions previously developed in the AMO community, e.g. for ion trapping~\cite{Pagano2018,Micke2019} or for Bose-Einstein condensation~\cite{Roux2008,Bernon2013}, is not possible. 

In this work, we have chosen to keep, whenever possible, the technical solutions adopted in our existing room-temperature setup, e.g. the use of in-vacuum high-NA aspheric lenses or that of a Zeeman slower as an atom source~\cite{Beguin2013PhD}. This has allowed us to focus mainly on the design of the cryogenic part. We base our design on the use of a closed-cycle cryostat using a pulse-tube refrigerator (PTR), with the technical constraint of using only UHV-compatible materials. However, to keep the design of our custom-made cryostat close to that of a commercial model~\cite{optidry}, we opt for a non-bakeable system, as the PTR cannot be baked out without being damaged (having a removable PTR to allow for bakeout of the rest of the system makes the design significantly  more involved). This tradeoff results in having a moderate vacuum in the room-temperature chamber, but as we shall see, cryopumping by the 4~K shield enclosing the atoms still results in long trapping lifetimes. 

\begin{figure*}[t]
\begin{center}
\includegraphics[width=0.9\textwidth]{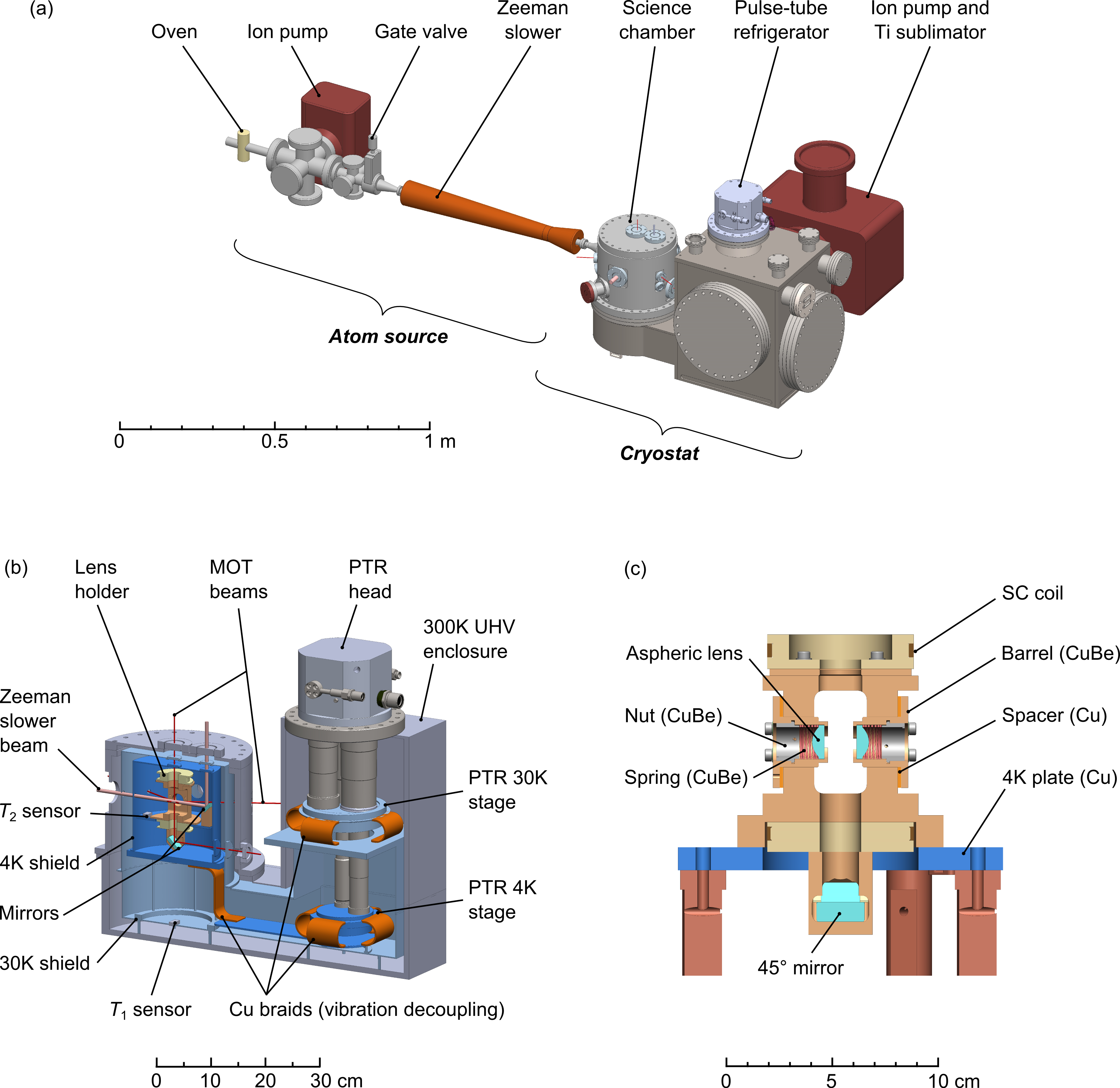}
\caption{Experimental setup. (a) Schematic rendering of the entire apparatus, comprising the atomic source and the cryostat. (b) Longitudinal cut of the cryostat, showing the pulse-tube refrigerator with its two stages at 30~K and 4~K, together with the corresponding thermal shields to which they are connected through vibration-decoupling copper braids. (c) Close-up cut of the lens-holder piece. }
\label{fig:setup}
\end{center}
\end{figure*}

Figure \ref{fig:setup}(a) shows a general view of the system. The cryostat is enclosed in a large stainless steel  vacuum chamber at 300~K that accommodates the PTR on one side, and a science chamber on the other side. An atomic source, comprising a rubidium oven followed by a Zeeman slower, is connected to the science chamber, and can be isolated from it using a gate valve actuated with a stepper-motor. The cryostat chamber is pumped using a 300~L$/$s ion pump (that includes a titanium sublimator), as well as with a Non-Evaporable Getter (NEG) cartdrige. 

A cut of the cryostat assembly is shown in Fig.\ref{fig:setup}(b). The two cooling stages of the PTR at 30~K and 4~K are thermally connected to nested, gold-plated copper radiation shields, that extend all the way to the science chamber. This connection is made using ultra-soft, high thermal conductivity copper braids for vibration decoupling. On the thermal shields, antireflection-coated 5-mm thick fused silica windows allow for optical access along all the needed directions~\cite{Footnote-windows}. The vibrational decoupling with copper braids is highly efficient: with the PTR in operation, we measure, along the three orthogonal directions, residual vibrations on the 4~K-baseplate below 10~nm (rms), the main frequency components being in the Hz range. 

The optical assembly for atom trapping is bolted on the 4K-baseplate, in the center of the science chamber, and comprises a beryllium-copper (CuBe) lens-holder and two  mirrors for beam-steering. The four MOT beams in the horizontal plane, as well as the tweezers beam along the  optical axis of the aspheric lenses, propagate in a straight line from outside the chamber, through a total of two vacuum viewports and four windows on the thermal shields, and exit the chamber on the other side. Three beams, on two axes (the Zeeman slower beam and the vertical MOT beams) are reflected inside the chamber on $45^\circ$ metallic mirrors held by CuBe supports. This allows (i) for the vertical MOT axis, to avoid having beams coming from below the chamber, which would make the construction of the cryostat quite involved, and (ii) for the Zeeman slower beam, to avoid having a cold window facing the atomic beam, where rubidium would accumulate, rendering it opaque. Two apertures with a diameter of 13~mm, one in each thermal shield, allow the atomic beam from the Zeeman slower to enter the trapping region. 

Figure~\ref{fig:setup}(c) shows a cut view of the lens mount. It is milled in a CuBe block; this choice of material is a trade-off to retain a good thermal conductivity while having better mechanical properties than copper~\cite{Ekin}. The two aspheric lenses (LightPath technologies, numerical aperture $0.5$, focal length 10~mm, working distance 7~mm) are mounted in CuBe barrels. To account for the differential thermal contraction between CuBe and glass upon cooling, the barrels were machined such that, at room temperature, their internal diameter exceeds the outer diameter of the lenses by 20~$\mu$m, resulting in a perfect match of diameters at 4~K. The flat face of the lens is pressed against a shoulder at the end of the barrel using  a CuBe spring and a nut to ensure the correct positioning of the lens at the end of the barrel. In a preliminary set of experiments we checked, using white-light illumination between crossed polarizers, that no stress-induced birefringence occurs in the lenses when cooling down the system. 

The two lenses are mounted with a spacing such that at 4~K, they are in an ideal $f-f$ configuration. Due to the thermal contraction of the CuBe lens holder, and to a lesser extent to that of the aspheric lenses, this means that at room temperature an incident collimated beam will focus at a finite distance, calculated to be $\sim 2.5$~m, after passing through both lenses. Using copper spacers between the barrels and the holder, with a thickness that we gradually reduce by lapping, the longitudinal positioning of the lenses is carefully adjusted until the proper spacing is obtained. When cooled down to 4~K, the system becomes almost afocal, as a collimated incident beam focuses at a distance  $>20$~m after the second lens. 

In view of future experiments with Rydberg atoms, the face of the lens facing the atoms is coated with a transparent but conductive layer of indium-tin oxyde (ITO), with a thickness of  120~nm (giving an overall transmission of  the lens of about 90\% at the tweezers wavelength of 830~nm). 

The lens holder also accommodates two independent superconducting coils, wound with 0.5~mm diameter NbTi wire, that can be used to produce the MOT magnetic field  gradient or a homogeneous bias field when switching from an anti-Helmholtz to a Helmholtz configuration. They are connected to the exterior of the cryostat via 0.6~mm diameter, kapton-insulated copper wire (the chosen diameter is a trade-off, that minimizes the heat conduction from room temperature to 4~K and the Joule heating in the wire, for the design current of 2~A~\cite{Ekin}). To minimize the effect of eddy currents when switching the magnetic field on and off, the coil form is made of CuBe which, unlike pure copper, has a moderate electrical conductivity even at cryogenic temperatures. In a preliminary experiment in a test cryostat at 4~K, we measured decay times of $\sim 1$~ms for the magnetic field; however, there, the copper thermal shields were quite remote from the coils. In the final configuration of the cryostat, when operating a MOT (see section~\ref{sec:mot}) and turning off the field, we have observed that the magnetic field experienced by the atoms fully settles after only $\sim 40$~ms, most likely due to the presence of pure copper parts (in particular the 4~K thermal shield) close to the coils. While this is not an issue for loading optical tweezers as we show below, one could improve on this in future designs by replacing some copper parts by CuBe ones when possible, and cutting out narrow slits in the shields at appropriate locations to break the paths of eddy currents. Finally, we have checked that for relevant repetition rates of current switching, eddy currents do not lead to any appreciable heating.

\subsection{Performance of the cryostat}

To operate the cryostat, we first evacuate the system with turbomolecular pumps, until we reach a residual pressure in the $10^{-8}$~mbar range. This pressure is   due to (i) the absence of bake-out of the setup, and (ii) the large number of elements under vacuum, especially those having a large surface-to-volume area, such as the copper braids, for which outgassing is very slow. We then switch on the PTR, and, within about 15 hours, the temperatures $T_1$ and $T_2$ measured by sensors on the 1st-stage (``30~K'') and on the 2nd stage (``4~K'') shields reach steady values. The pressure in the chamber, as measured by the 300~L$/$s ion pump current, is then around $4\times 10^{-10}$~mbar \cite{VacuumLeak1}. 

Warming up the system to room temperature takes about 100~h when keeping the chamber under vacuum; if needed, faster cycling times could be achieved by flushing the chamber with dry nitrogen to enhance heat exchange. 

We characterized the performance of the cryostat in a series of preliminary experiments in various configurations, which allowed us to evaluate its response to the various heat loads it is subjected to in operation. We first cooled down the system in a configuration minimizing the heat load (no wiring for the SC coils, windows in the thermal shields were replaced by gold-plated copper blanks, openings for the atomic beam were sealed) and measured $T_1=30.1$~K and $T_2=3.2$~K, which gives the base temperature the system can reach. By applying controlled power to heaters located on the 4~K plate, we measure a temperature increase of around 4~K$/$W, which gives an estimate of the acceptable heat load. In a second configuration where the fused silica windows are mounted on the thermal shields and the apertures for the atomic beam are open, the measured temperature is barely affected, showing that most of the black-body radiation is effectively blocked by the windows. In the final configuration, the SC coils are connected using their four 0.6~mm diameter wires; the measured temperature (without any current flowing in the coils) is then $T_2=4.2$~K, consistent with the heat load due to heat conduction along the wires. 

Finally, we test the cryostat performance in the presence of the two extra heat load sources that appear when trapping atoms, namely laser light for the tweezers array, and current flowing through the coils. Concerning laser power, due to the ITO coating on the lenses, a significant part (about 20\,\%) of the light at 830~nm is absorbed or reflected by the pair of lenses and does not exit the cryostat; part of it is thus a direct heat load for the 4~K environment. For an incident power of 1~W (enough to generate about 500 optical tweezers), we measured a temperature increase of the lens holder by about 1~K.  Concerning the operation of the coils, we observe a slight temperature increase (0.1~K for 1~A) when we run a current through them. For small currents, up to $1.7$~A (corresponding to a MOT gradient 7.3~G$/$cm), we attribute this to Joule heating of the (non-SC) wires connecting the coils to the room-temperature connectors. Beyond this value, we observe a jump in the coil resistance, indicating that they partially reach a temperature above the NbTi critical temperature of 9.2~K  and transition to the normal state, most likely because the thermal contact between the kapton-insulated SC wire and the CuBe coil form is not sufficient for proper thermalization. Then, the temperature increase is steeper, with the lens holder sensor reaching a temperature of 5.4~K when the current is 2.5~A. This is more than enough for operating a MOT in order to load the tweezers array, as we discuss in section~\ref{sec:mot}.

\section{Single-atom trapping in arrays of optical tweezers}

\subsection{Magneto-optical trap}
\label{sec:mot}

\begin{figure}[t]
\begin{center}
\includegraphics[width=80mm]{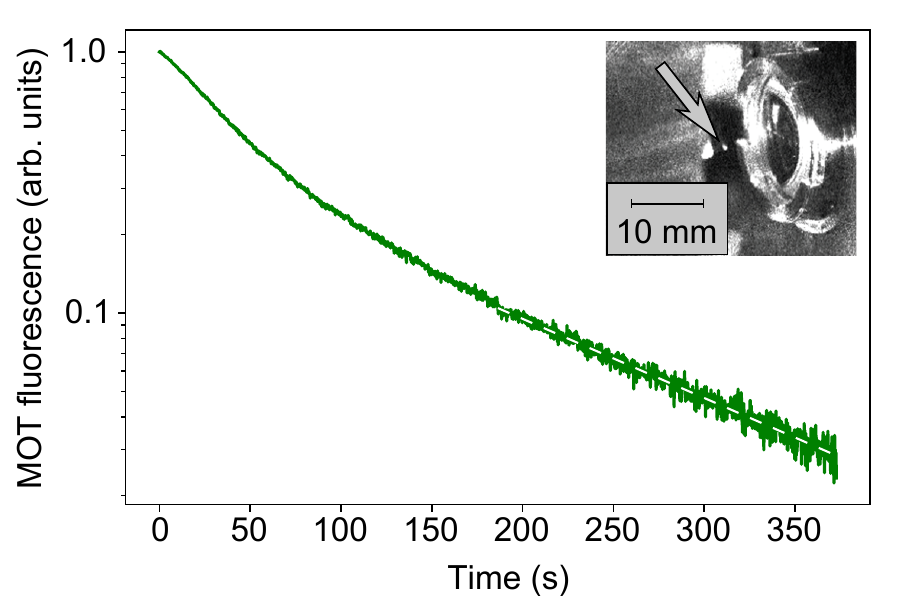}
\end{center}
\caption{Fluorescence decay of the MOT. An exponential decay fit at long times (white dashed line) gives a $1/e$ decay time of about 140~s. The inset shows the MOT cloud (arrow) facing one of the aspheric lenses.}
\label{fig:MOT}
\end{figure}

We now describe the operation of the setup for atom trapping, starting with the realization of a $^{87}$Rb magneto-optical trap (MOT). To do so, we operate the rubidium oven typically at $100^\circ$C. The resulting atomic beam is slowed down via the Zeeman slower and loads a magneto-optical trap in the science chamber. The MOT uses six counter-propagating lasers beams with a $1/e^2$ radius of 1.7~mm and a power of 1~mW each, detuned by $-4.5\Gamma$ from the $F=2 \rightarrow F'=3$ transition of the $D_2$ line (whose natural linewidth is $\Gamma=2\pi\times6$~MHz). Repumping light is combined with these six beams, with a power 0.1~mW per beam; it is resonant with the $F=1 \rightarrow F'=2$ transition of the $D_1$ line. The typical magnetic field gradient used for MOT loading is 6~G$/$cm.

After loading the MOT for typically 500~ms, we turn off the Zeeman slower beam and close the gate valve  to stop any further loading of the magneto-optical trap. The decay of the MOT fluorescence, measured with a CCD camera, is shown in Fig.~\ref{fig:MOT}. At short times, the MOT decays relatively quickly, due to a combination of (i) light-assisted collisions in the dense central region of the cloud and (ii) the escape of atoms from the outer regions of the MOT, where beam intensities are not perfectly balanced (making this initial decay quite sensitive to the alignment of the MOT beams).  At long times, the fluorescence decay is exponential, with a $1/e$ lifetime of about 140~s, much less sensitive to beam alignment. Such a MOT lifetime is typical of vacuum systems with pressures in the low $10^{-12}$~mbar range, showing the dramatic effect of cryopumping by the 4~K surfaces surrounding the atoms, despite the relatively low vacuum in the room-temperature chamber. This measured lifetime gives a lower bound on the vacuum-limited lifetime we can expect for atoms in optical tweezers \cite{VacuumLeak2}.

\subsection{Arrays of optical tweezers}

\begin{figure}[b]
\begin{center}
\includegraphics[width=85mm]{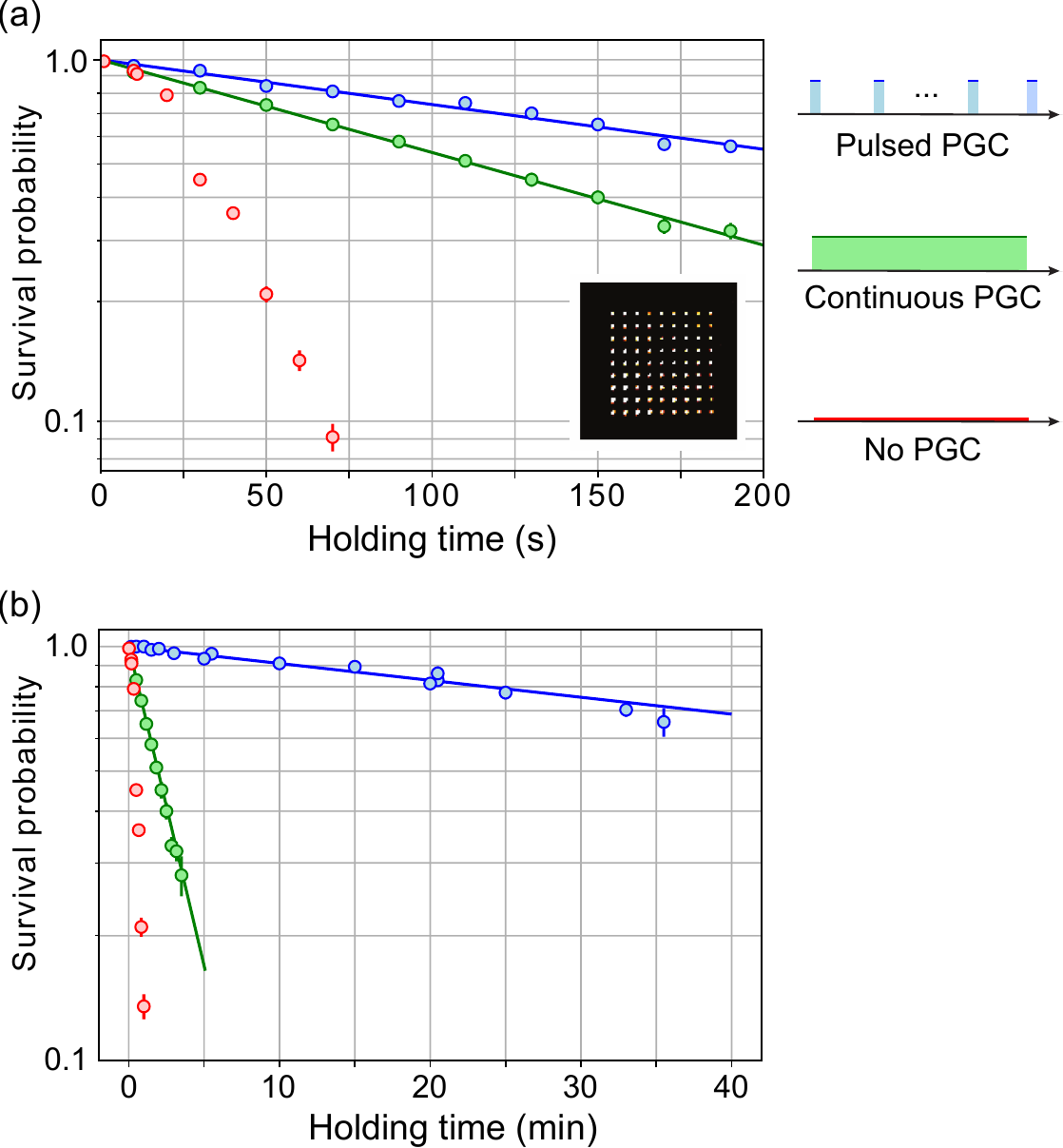}
\end{center}
\caption{(a) Survival probability of a single atom as a function of the time it is held in an optical tweezers, either without PGC (red), with continuous PGC (green), and with a 15~ms pulse of PGC every 10~s (blue). Exponential decay fits (solid lines) give $1/e$ decay times of about 162~s for continuous PGC, and of 335~s for pulsed PGC. Inset: average fluorescence image of the $9\times9$ tweezers array, with a spacing of $10\,\mu$m between adjacent microtraps. (b) Lifetime measurement for pulsed PGC (blue) after improvement of the vacuum \cite{VacuumLeak2}, note the change in the horizontal scale. The $1/e$ decay time is now 6050~s (solid line). For comparison, the no-PGC and continuous PGC curves from panel (a) are plotted again.}
\label{fig:Lifetime}
\end{figure}

We then study the loading of single atoms into optical tweezers, which are created using light at 830~nm. Using a spatial light modulator, we create arbitrary tweezer arrays in the focal plane of the aspheric lens~\cite{Nogrette2014}. For the work reported here, we use a $9\times9$ square array (see the average fluorescence image in the inset of Fig. \ref{fig:Lifetime}). The fluorescence emitted by trapped atoms is collected, using the same aspheric lenses, on an EMCCD camera with a typical exposure time of 50~ms. We observe that despite the large number of optical surfaces the beams go through and the high reflectivity of the gold-plated thermal shields, stray light is barely higher than in our room-temperature setup, and does not affect significantly the detection of single atoms.

For a power of 3~mW per tweezers, we measure, using parametric heating, an axial (radial) trapping frequency 8~kHz (70~kHz). The trap depth is $U_0/k_{\rm B}\simeq 0.8$~mK. Using a release and recapture method \cite{Tuchendler2008}, we measure the atomic temperature in the tweezers to be around 50~$\mu$K after the atoms have been cooled for 50~ms by polarization gradient cooling (PGC) with a detuning of $-4.5\Gamma$. We further cool the atoms down to 20~$\mu$K, using a $-10.5\Gamma$ detuned light pulse of 40~ms.

To measure the time evolution of the probability to keep an atom in the optical tweezers, we record a first fluorescence image to identify the traps initially containing atoms, we then wait for a time $t_{\rm hold}$, and we finally take a second image to identify the remaining atoms. 

Without any cooling light during the hold time, half of the atoms are lost after about 30~s, and the decay of the recapture probability with time is non-exponential (see red dots in Fig.~\ref{fig:Lifetime}). This is explained by a linear heating rate, that we measured in a separate experiment to be of about 8~$\mu$K$/$s, originating from off-resonant scattering of the 830~nm trapping light. An obvious way to mitigate this heating is to leave PGC on during the hold time. With a detuning of $-10.5\Gamma$, the recapture probability is  then increased drastically, giving an exponential decay with a $1/e$ decay time of $162$~s (green dots). However, a careful inspection of the second image shows that occasionally, an initially empty trap is occupied in the final image. A more detailed analysis (see Appendix~\ref{app:a}) shows that from time to time, some atoms that are expelled from a trap via collisions with background-gas molecules are still slow enough to be recaptured in the optical molasses, and then reloaded in another optical tweezers, either giving rise to a trap loading (if this other trap was initially empty) or to the correlated loss of two atoms (if the other trap was already occupied). This suggests that the trap lifetime can be further increased.  

To do so, we pulse the PGC cooling light, sending a 15~ms PGC pulse at $-10.5\Gamma$ every 10~s. These timings fulfill the following conditions: the PGC pulse is long enough to fully cool again the atom, and repeated often enough such that the increase in temperature induced by the 830~nm light over the 10~s remains well below the trap depth. At the same time, the overall duty cycle $\eta=0.15\%$ is very small, such that the probability of correlated atom loss, now multiplied by $\eta$, becomes entirely negligible. In these conditions, we measure a background-collision limited lifetime of 335~s (blue), i.e. an improvement by a factor $\sim16$ as compared to our room-temperature setup \cite{VacuumLeak2}. Finally, after improving the vacuum, we repeat the pulsed-PGC lifetime measurement and obtain a $1/e$ lifetime of 6050~s as shown in Fig. \ref{fig:Lifetime}(b). The pulsed PGC cooling is entirely compatible with atom-by-atom rearrangement, meaning that we can benefit from this lifetime increase for assembling large arrays.  

\section*{Conclusion}

In this work, we have demonstrated, using a relatively simple setup, the trapping of single atoms in arrays of optical tweezers in a 4~K environment, with long lifetimes of over 6000~s, that open exciting prospects. We now discuss possible ways to improve the performance in the future. 

Using the same setup, the next step will consist in realizing large rearranged arrays with hundreds of single atoms. Defect-free arrays of $\sim800$-atoms seem to be within reach in our setup (the necessary 1,600 optical tweezers still correspond, for a trapping wavelength of 830~nm, to an acceptable heat load for the cryostat). In the current stage, with ITO coating on the aspheric lenses and appropriate anti-reflection coatings on the windows, the setup is compatible with Rydberg excitation, albeit without the possibility of electric field control. A direct measurement of Rydberg level lifetimes using a ponderomotive bottle-beam trap~\cite{Barredo2020} would be interesting, to check the Rydberg state lifetime increase due to the suppression of BBR-induced transitions. Adding a set of electrodes on the lens holder, and possibly a microwave antenna for coherent manipulation in the Rydberg manifold, will be a relatively simple upgrade of the current setup.

To improve the residual pressure even further, the ultimate step would be to make the system bakeable. For that, the design needs to use a removable PTR, which requires to use radiators in a chamber filled with buffer gas as the vibration-decoupling heat exchanger, in place of the copper braids used here. Another possible improvement would be to maximize the cryopumping efficiency using porous materials such as activated charcoal. Such a setup, although more involved than the one used in the current work, is perfectly realistic. Cryogenic setups will certainly allow reaching 1000-atom scale tweezers arrays, and maybe even more if combined with techniques~\cite{Sompet2013,Lester2015,Brown2019,Aliyu2021} that allow for an initial loading efficiency of the array way above 50\%, thus reducing both the assembly time and the required trapping laser power.

\begin{acknowledgments}
We thank Eric Magnan and Sam~R. Cohen for contributions in the early stages of the experiment, as well as Michel Brune, Jean-Michel Raimond, and Cl\'ement Sayrin for useful discussions, and Igor Ferrier-Barbut for a careful reading of the manuscript. Mohammed Sharazi and Franck Ferreyrol from My Cryo Firm were involved in the design and construction of the custom-made cryostat. KNS acknowledges funding from the Studienstiftung des deutschen Volkes. SP is partially supported by the Erasmus+ program of the EU. DB acknowledges support from the Ram\'{o}n y Cajal program (RYC2018-025348-I). This project has received funding from the R\'egion \^Ile-de-France through the DIM SIRTEQ (project CARAQUES) and from the European Union's Horizon 2020 research and innovation program under grant agreement no. 817482 (PASQuanS). 
\end{acknowledgments}

\appendix
\section{Correlated loss and recapture under continuous polarization-gradient cooling}
\label{app:a}

The fact that the measured trapping lifetime for an atom in an optical tweezers is reduced under continuous PGC conditions can arise from two different effects, whose relative importance depends on the experimental parameters. 

The first effect is simply that when the PGC beams are always on, a steady-state, very dilute cloud of laser-cooled atoms (loaded either from slow atoms from the source, either from a residual Rb pressure in the chamber), always surrounds the tweezers array; this yields occasional loading of single atoms in a tweezers that was already occupied, resulting in the loss of both atoms. In the present case, this effect should be negligible, as the atom source is mechanically blocked by the stepper-motor-actuated valve, and the residual Rb pressure in the 4~K environment is extremely small.

\begin{figure}[b]
\begin{center}
\includegraphics[width=85mm]{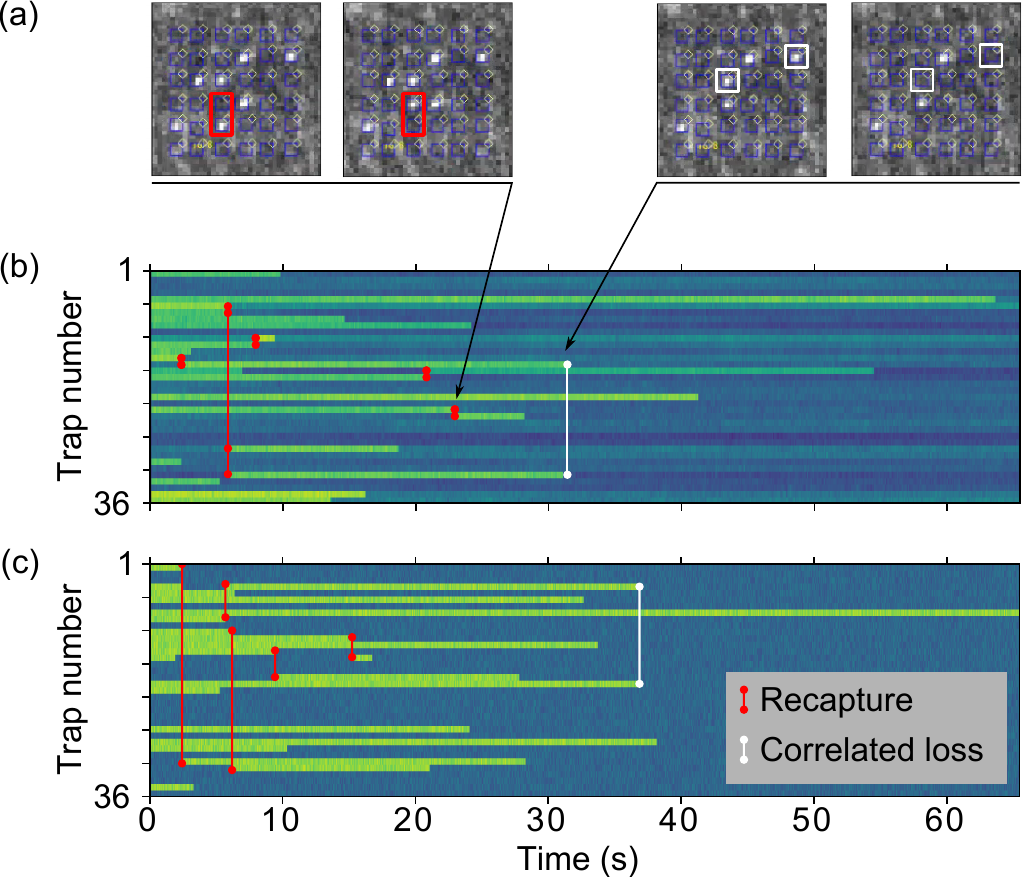}
\end{center}
\caption{(a) Examples of fluorescence images of a $6\times6$ array showing, between successive frames, either the loss of an atom accompanied by the loading of a previously empty trap (left), or the ``simultaneous'' loss of two atoms (right). (b) Experimental fluorescence traces showing the evolution of the array occupancy as a function of time, with correlated losses and recapture events highlighted in white and red, respectively. (c) Result of a Monte-Carlo simulation of the simple model discussed in the text, showing the same qualitative behavior as the experimental traces in (b).}
\label{fig:appendix}
\end{figure}

The second effect is the following. The energy that is imparted to a trapped Rb atom by a molecule from the residual gas in the vacuum chamber (consisting mostly of ${\rm H}_2$ molecules, as most other species are extremely well cryo-pumped by the 4~K walls) can be small enough that the Rb atom, while expelled from the $\sim~1$~mK-deep optical tweezers, is still captured in the optical molasses~\cite{CollisionEnergyNote}. This atom can then be very quickly loaded in another trap of the array, that was either empty or loaded. In the first case, two successive frames of the camera that monitors the fluorescence of the array will show the same number of trapped atoms, but with one trap having lost its atom, and another one being suddenly loaded (Fig.~\ref{fig:appendix}(a), left). In the second case, the second frame will show two atoms less than the first one. (Fig.~\ref{fig:appendix}(a), right).

Analyzing the successive images acquired during continuous PGC (but taken with a detuning of $-4.5\Gamma$ to get relatively bright fluorescence images) shows that several of those correlated losses and recapture events can be identified during the full decay of the array, and that they significantly contribute to the trapping lifetime. A typical example of such an analysis is shown in Fig.~\ref{fig:appendix}(b). Figure~\ref{fig:appendix}(c) shows the result of a very simple stochastic modeling of the process. At each time step, corresponding to an imaging frame, each atom in a filled trap $i$ has a probability $p_{\rm coll}$ to undergo a collision with the background gas; if a collision does occur, it leaves the trap but has a probability $p_{\rm rec}$ to be recaptured in \emph{any} trap $j$ of the array (including $i$), chosen randomly, giving rise to either a recapture or to correlated loss. We find that values around  $p_{\rm rec}\sim 0.2$ reproduce qualitatively the main features of the experimental traces. 

A quantitative investigation of the dependence of $p_{\rm rec}$ on various parameters (temperature of the environment, parameters of the PGC) is beyond the scope of this paper, but could be an interesting extension of the present work.  

\begin{thebibliography}{99}
\bibitem{NoteSQRT}
If the single-atom lifetime is $\tau$, the lifetime of an $N$-atom configuration is $\tau/N$. As the rearrangement time scales roughly linearly with $N$, an improvement by a factor $\alpha$ of $\tau$ results in an increase of the maximal atom number that one can rearrange during the lifetime of a configuration by only  $\sqrt{\alpha}$.
\bibitem{Footnote-windows}
In order to avoid stress-induced birefringence upon cooling down, the windows are not rigidly mounded on the thermal shields, but held by a retaining ring which is simply pressed against the shields using beryllium-copper strips of fingers.
\bibitem{CollisionEnergyNote}
At $T=4$~K, a $\rm H_2$ molecule has a typical thermal velocity of $\sqrt{3k_{B}T/m_{\rm H_2}}\simeq 220\,{\rm m/s}$. Even in the worst case scenario of a head-on collision, a molecule with this velocity will give to the heavy $^{87}$Rb atom (initially almost at rest in its tweezers) a velocity of only $\sim 10\;{\rm m/s}$ (and obviously way smaller for grazing-incidence collisions). Thus, a very significant fraction of collisions with the residual gas gives Rb atoms with a velocity within the capture range of the optical molasses. At room temperature, the effect would be way less frequent, as the thermal velocity of ${\rm H}_2$ molecules is higher by a factor of almost~9.
\bibitem{VacuumLeak1}
The pressure values stated here were measured after an improvement of the room-temperature vacuum. Indeed, after  acceptance of the paper, we noticed a leak in the titanium sublimation pump. Repairing the issue improved the room-temperature vacuum, as measured by the ion pump, by two orders of magnitude.
\bibitem{VacuumLeak2}
Here, the stated lifetime values were measured before the vacuum was improved. At cryogenic temperatures, the improvements resulted in a 20-fold increase of the measured single-atom lifetimes in the optical tweezers of over 6000~s (see Figure 3b).
\end{thebibliography}
\end{document}